\documentstyle[11pt,psfig]{article}
\newcommand{\be}{\begin{equation}}
\newcommand{\ee}{\end{equation}}
\newcommand{\bea}{\begin{eqnarray}}
\newcommand{\eea}{\end{eqnarray}}
\newcommand{\beas}{\begin{eqnarray*}}
\newcommand{\eeas}{\end{eqnarray*}}
\newcommand{\avg}[1]{\left\langle{#1}\right\rangle} 
 
\newcommand{\ovl}[1]{\overline{#1}}

\begin{document}

\title{Statistical mechanics of asset markets with private information}
\author{Johannes Berg$^1$, Matteo Marsili$^2$,\\ Aldo Rustichini$^3$, 
\& Riccardo Zecchina$^1$\\
~~~~~\\
{\em 1 - Abdus Salam International Centre for Theoretical Physics,}\\ 
{\em 34100 Trieste, Italy}\\
{\em 2 - Istituto Nazionale per la Fisica della Materia (INFM),}\\
{\em Unit\'a Trieste-SISSA, 34014 Trieste, Italy}\\
{\em 3 - Dept. of Economics, University of Minnesota,}\\
{\em Minneapolis, MN, 55455, USA}
}
\maketitle
 
\begin{abstract}
Traders in a market typically have widely different, private
information on the return of an asset. The equilibrium price of the
asset may reflect this information more accurately if the number of
traders is large enough compared to the number of the states of the
world that determine the return of the asset. We study the transition
from markets where prices do not reflect the information accurately
into markets where it does. In competitive markets, this transition
takes place suddenly, at a critical value of the ratio between number
of states and number of traders. The Nash equilibrium market behaves
quite differently from a competitive market even in the limit of large
economies.
\end{abstract}

\newpage
\section{Introduction}

The fundamental hypothesis, which is typically adopted in financial
economics, is that markets are efficient. This general hypothesis has
been recently the object of a detailed, critical evaluation (see for
instance \cite{CE}), nevertheless it remains an extremely useful
benchmark.  Many possible definitions of this hypothesis have been
given. We shall adopt here the one suggested by Malkiel
\cite{efficiency}: A market is efficient with respect to an
information set if the public revelation of that information would not
change the prices of the securities.  Loosely speaking, this means
that the information is incorporated into, or reflected by, prices.

The efficiency hypotheses may also differ for the degree, or {\it form} of efficiency. The one we are going to consider here is {\it strong efficiency}: The information set includes the information available to any of the participants in the market, including private information. 

In order to address the issue of efficiency, it is fundamental to
understand how precisely information scattered across different agents
is aggregated into the prices of the assets that are traded in the
market. Two fundamental insights of economic theory go a long way
towards addressing this question.

The first insight is that the asymmetric information of different
traders may cause inefficiency of the equilibrium \cite{Ak}, given the
strategic incentive of each agent not to reveal the information he
has. The second insight is that in {\it large} markets the distortion
produced by the private incentives and asymmetric information may
vanish, because the benefit from the distortion, as well as the
ability to distort, for any single agent may be infinitesimally
small. This second insight makes the concept of {\it market
efficiency} interesting and fruitful.

Here we address a specific aspect of the general question of
efficiency of financial markets. How precisely does the size of the
market affect efficiency? Or more precisely, how does the relative
size of the market compared to the size of the uncertainty that agents
are facing affect efficiency, when information available is private?
The ``size of the market'' is an unambiguous concept, and is measured
by the number of agents. The ``size of the uncertainty'' is less
commonly used: It is defined here to be the number of possible states
of nature affecting the return of the asset. We address this question
in the context of a simple market, where the price of the asset is the
outcome of market clearing between a total monetary demand and a fixed
supply.

An alternative method, which has been frequently adopted in recent
years, is the formulation of a precise, detailed description of the
price formation process. For instance, the literature of market
micro-structure analyzes the process of price formation in concrete
markets, with a limited number of traders, and a well specified,
usually sequential series of moves that ultimately determine the price
of the asset. The emphasis is on the strategic aspects of information
revelation through trade, and the role that institutional details of
the market considered play in the process. 

This method has provided useful insights in the working of concrete
markets \cite{OH}. But as always when one adopts very specific setups,
the results depend in a critical way on the details of the extensive
form game. Instead, we consider here a very simple market. There is
only one asset, with a return that depends on the state of nature.
Traders observe an imperfect signal of the realization of the state,
and can then decide how much they invest in the asset, {\it before}
the price of the asset is announced. The price of the asset is then
determined to clear the market. Note that the agents must decide their
investment before they know the price for the period, and the price is
determined to equate the total demand equal to the total amount
invested, to a fixed supply. So this model is very similar to the
classic model of Shapley and Shubik \cite{ShapleyShubik}.

If agents had a complete knowledge of the state, competitive prices
would equal returns in any state. This means that prices would reflect
all the information on the state of nature, i.e.  that the market
would be efficient. When agents have only partial information,
competitive prices may fail to equal returns and the market may not be
efficient. It will turn out that the fundamental element for the
efficiency is the size of the market compared to the size of the
uncertainty. Therefore we are going to introduce as crucial parameter
the {\it relative} size of the market, that is the number of agents in
the market divided by the number of the states of nature. Our first
main result is that the competitive equilibrium of the market becomes
efficient suddenly, once a threshold in relative number of agents and
events is crossed. 

The second question we analyze is the size of the difference between
competitive and Nash equilibrium prices. This difference, and in
particular its asymptotic behavior as the number of agents tends to
infinity has already been studied, although in different
setups. Particularly relevant from the point of view of our research
are the results that deal with economies with private information. The
typical result is that the difference between Nash and competitive
equilibrium vanishes as the number of agents grows. The reason for
this is clear: with a large number $N$ of agents the effect of the action
of each agent on the aggregate outcome becomes negligible, and
therefore the incentive to behave differently than an agent in a
competitive economy vanishes. The interesting object of study
therefore is the rate of convergence to zero of this difference. This
rate depends on the specific game, and concrete examples yielding
different rates have been found. Two examples may be considered here
as an illustration. In the Gul and Postlewaite paper \cite{GP} the
rate of convergence is $\frac{1}{\sqrt{N}}$. But in Rustichini,
Satterthwaite and Williams \cite{RSW} the rate is much faster, and
is equal to $\frac{1}{N}$.

In our model an additional feature enters the analysis: the number of
agents {\it and} the size of the uncertainty, modeled by the number of
states of nature, are growing together.  We are going to show that the
relative difference of competitive versus Nash equilibrium prices may
be non vanishing in the limits of large markets. Perhaps more
surprisingly, the speed of convergence depends on the amount of
information available, and in a discontinuous way.

Our results are based on tools and ideas of statistical mechanics of
disordered systems \cite{MPV}. These same ideas have recently proven
quite useful in the study of systems of heterogeneous interacting
agents \cite{CMZ}.

\section{The model}
The model is a simple market with one asset, and many traders. The return of the asset is determined in each period by the draw of a state of nature. Agents do not know the state, but observe an imperfect signal on it. After they observe the signal they have to decide the amount they invest in that period. The total amount invested, divided by the total supply of the asset, which is fixed, determines the market-clearing price. Each agent gets then the return of the asset for each unit he owns, and the market goes to the next period. 

The return in each state, as well as the type of signal available to each agent, are determined once and for all before the initial period. And now we proceed to a more detailed exposition. 

\subsection*{Asset, returns, and agents}
A single asset is being traded over an infinite number of periods in a market with a set $I$ of $N$ traders. There are $N$ units of asset available. The asset has a monetary return paid at the end of each period. This return is different in each period, depends on the state of nature for that period, $\omega \in \Omega$, and is denoted by $R^{\omega}$. The state of nature is determined in each period, independently, according to the uniform distribution on $\Omega$, which is assumed to be a finite set\footnote{We shall use
$\Omega$ also to denote the number of elements of $\Omega$.}.  The value of the return $R^{\omega}$ for each state $\omega$ is drawn at random before the first period, and does not change afterwards.  Returns thus only change because the state of nature changes.

Traders do not observe the state directly, but have a signal on the state according to some fixed private information structure, which is determined at the initial time and remains fixed. More precisely,  a signal is a function from the state $\omega\in\Omega$ to
a signal space, which for simplicity we assume to be $M \equiv \{-, +\}$. The signal observed by trader $i$ if state is $\omega$ is $k_i^
\omega$.  The information structure available to $i$ is the vector $(k_i^\omega)_{\omega \in \Omega}$. This structure is determined, by
setting $k_i^\omega =+1$ or $-1$ with equal probability, independently across traders $i$ {\it and} states $\omega$.

How revealing the information provided by $k_{i}$ is depends, at each
state $\omega$, on the entire realization of the vector, as well as on
$R^\omega$. If, for example $k_i^\omega=+$ for all $\omega$ such that
$R^\omega>\bar R$ (and $k_i^\omega=-$ otherwise), then agent $i$ will
know for sure when the return is higher than $\bar R$. In the random
economy we consider, this is a very unlikely situation, because
$R^\omega$ and $k_i^\omega$ are drawn independently. Still for agent
$i$ the distribution of returns conditional on the signal he receives,
will depend to some extent on the signal. Note that an agent who 
knew simultaneously the partial information of all agents would be
able to know the state $\omega$, with probability one, for $N\to\infty$. Indeed the probability that there are two states $\omega$ and $\omega'$ with different returns and that no agent can distinguish them is well approximated by $\Omega(\Omega-1)2^{-(N+1)}$, which vanishes for $N\to\infty$, even if $\Omega$ grows proportionally to $N$ as we shall assume later.

So if the information available to each single agent happened to be revealed to all, then prices would be equal to return, for all states but a subset of states with measure tending to zero. This is what the strong efficiency hypothesis would require in our model.

\subsection*{The market}
At the beginning of each period each trader decides to ``invest'' a
monetary amount $z_i^m$ in the asset,  depending on the signal $m=k_i^\omega$ which agent $i$ receives at that time. The total amount invested by all the agents is the demand of the asset, and the supply is fixed to $N$. The market clearing condition determines the price $p^{\omega}$ for the period, according to:

\begin{equation}
\label{eq:mcc}
\sum_{i\in I}\sum_{m\in M} z^m_i\delta_{k_i^\omega,m} = p^{\omega} N
\end{equation}
where $\delta_{i,j}$ is the Kronecker delta. When agents decide their investment $z_i^m$, they do not know the price at which they will buy the asset. Note that the price depends on the state since the amount invested by each agent depends on the state \cite{ShapleyShubik}.
At the end of the period, each unit of asset pays a monetary amount $R^\omega$. If agent $i$ has invested $z_i^m$ units of money, he 
will hold $z_i^m/p^\omega$ units of asset, so his payoff will be $z_i^m(\frac{R^{\omega}}{p^{\omega}}-1)$. 

The expected payoff of agent $i$ is described by a function $u_i$ of the entire vector of investments, as follows:
\begin{equation}
\label{eq:netr}
u_i(z_i, z_{-i})
=\frac{1}{\Omega}\sum_{\omega \in \Omega} 
\sum_{m\in M}\delta_{k_i^\omega, m} z^m_i
\left( \frac{R^{\omega}}{p^{\omega}}-1\right)=
\sum_{m\in M}\overline{\delta_{k_i, m} z^m_i
\left(\frac{R}{p}-1\right)},
\end{equation}
where for any real valued function $f$ defined on $\Omega$,
\[
\overline{f} \equiv \frac{1}{\Omega}
\sum_{\omega \in \Omega} f^{\omega}
\] 
and $z_{-i}$ denotes the strategy vector of all players but $i$. The goal of each agent $i$ is to maximize the expected value of $u_i$. 

Agents can, if they want, invest an infinite amount in a period. For simplicity we exclude the possibility that they can {\it sell} an infinite amount, since this would complicate the market clearing condition if we insist that the price of the asset has to be positive. The results of this paper hold if we first require that investment are less than a maximum amount, and then take the limit in which this bound tends to infinity.

\subsection*{Large markets}
We shall be interested in the behavior of the market in the limit when
$N\to\infty$. For the limit to be non-trivial limit, we need to discuss how different parameters of the model behave in the limit\footnote{We write as usual $x_n = O(y_n)$ for two sequences $\{x_n\}$ and
$\{y_n\}$ if for two positive constants $c$ and $C$, $c \leq
\frac{x_n}{y_n} \leq C$ for all $n$ sufficiently large. Also we write 
$\simeq$ for relations which hold almost surely as $N\to\infty$.}.

If returns are on average larger than prices, agents will tend to
invest more. This in turn drives prices up towards returns. We then
expect that average price $\overline{p}$ will converge to the average
return $\overline{R}$ and that the fluctuations of prices with
$\omega$ will adjust to those of $R^\omega$. The information structure
however constrains the fluctuations of prices. A simple
argument shows that price fluctuations are small:  Note that one
can write
\[
z_i^m=\frac{z_i^++z_i^-}{2}+m\frac{z_i^+-z_i^-}{2}
\]
for $m=\pm 1$.  Then for $\Omega$ large, 
$\overline{k_i}\simeq 0$ and $\overline{k_i \;k_j}\simeq
\delta_{i,j}$ and one easily finds that 
\[
\overline{p} \simeq \frac{1}{N}\sum_{i\in I} \frac{z_i^{+} + z_i^-}{2}
\]
and
\be
\overline{(p-\overline{p})^2}\simeq 
\frac{1}{N^2}
\sum_{i\in I}(\frac{z_i^{-} - z_i^-}{2})^2\propto \frac{1}{N}.
\label{pfluct}
\ee 

A non-trivial behavior is then expected when returns have fluctuations of a comparable size. Therefore we assume that returns have the specific form:
\begin{equation}
\label{eq:ret}
R^{\omega} = \ovl{R} + \frac{\tilde{R}^{\omega}}{\sqrt{N}}
\end{equation}
where $\ovl{R}>0$ and $\tilde R^\omega$ is drawn from a Gaussian
distribution with zero mean and variance $\sigma^2$. 

In particular we are interested in the transition to an efficient
market, where information about returns is incorporated into prices,
i.e. $p^\omega=R^\omega$. This is a set of $\Omega$ equations in
$2N$ unknown $z_i^m$. In order to allow for the possibility of
observing the transition from an inefficient to an efficient market,
we need to take a number of states $\Omega$ which is proportional to
$N$. Hence we define $\alpha=\Omega/N$. 

The key quantity we shall be interested in is the distance between 
vectors in $R^{\Omega}$, such as prices and returns, which
is defined as usual 
\[
|x-y|=\sqrt{\sum_{\omega\in\Omega}(x^\omega-y^\omega)^2}, 
\]
for all $x,y\in R^{\Omega}$. 

Finally note that the parameter $\ovl{R}$ can be set to $1$, without
loss of generality, by a suitable choice of the units of $z_i^m$. A
value $R\neq 1$ can be reintroduced at any stage of the calculation by
means of dimensional analysis. This leaves us with only two control
parameters $\alpha$ and $\sigma$ in the limit $N\to\infty$. Note now
that, with our choices of parameters, the distance of returns from
their average is finite: $|R-1|^2\simeq \alpha\sigma^2$.

Of course real markets have a finite, but large, number $N$ of agents
and finite, but small, fluctuations of returns. As we shall see
numerical simulations of finite, but large, economies are in perfect
agreement with the results for $N\to\infty$.

\subsection*{Competitive equilibrium and Nash equilibrium}
We are going to consider two equilibrium concepts for our market, competitive equilibrium and Nash equilibrium. Both give a description of what each agent does as a function of the signal observed. In the competitive equilibrium, each agent ignores the effect he has on the price. In the Nash equilibrium traders take their effect on prices into account. Here are the more precise definitions. 

A {\it competitive equilibrium} is a vector $(z^m_i)_{ i \in I, m \in M}$ such that:
\begin{enumerate}
\item for every $i$ and $m$, 
\[
z^m_i \in 
\mbox{ arg}\max_{x \geq 0}
x \overline{\delta_{k_i,m}
\left( \frac{R}{p}-1\right)}
\]
\item market clears at $p^\omega$, i.e. Eq. (\ref{eq:mcc}) holds.
\end{enumerate}
Note that if the expected net return $\overline{\delta_{k_i,m} \left(
\frac{R}{p}-1\right)}$ is positive then the trader invests an infinite
amount when receiving signal $m$.

A {\it Nash Equilibrium} is a vector $(z^m_i)_{ i \in I, m \in M}$ 
such that:
\begin{enumerate}
\item for every $i$ and $m$, 
\[
z^m_i \in 
\mbox{ arg}\max_{x \geq 0}
x \sum_{\omega \in \Omega} 
\delta_{k_i \omega} (m) 
\left( \frac{R^{\omega}}{p_{-i}^{\omega}+x/N}-1\right)
\]
where $p_{-i}^{\omega}=p^\omega-\sum_{m\in M}
\delta_{k_i^\omega,m} z_i^m/N$
is the contribution of all other agents to the price.
\item market clears at $p^\omega$, i.e. Eq. (\ref{eq:mcc}) holds.
\end{enumerate}
The difference between a competitive equilibrium and a Nash equilibrium 
usually vanishes as $N \rightarrow \infty$. 
It is easy to check this when agents have no information
(for instance, when $k_i^\omega=+1$, $\forall i\in I,~\omega\in\Omega$). 
In this case
price cannot depend on $\omega$. The competitive
equilibrium price, $p_{C}$ is equal to the expected return:
\[
p_{C} = \overline{R}
\]
while the Nash equilibrium prices are 
\[
p_{N} = \left(1 - \frac{1}{N}\right)
\overline{R}
\]
In this case the distance in $R^\Omega$ 
between the two prices vanishes as $1/\sqrt{N}$,
for $N\to\infty$. Note that the difference $p_C-p_N\propto 1/N$
is much smaller than 
fluctuations of returns, which vanish as $1/\sqrt{N}$. We are going to
see that, for a random information structure $k_i^\omega$, the
difference $p_C^\omega-p_N^\omega$ is generally of order 
$\frac{1}{\sqrt{N}}$. In particular it
is of the same order of the fluctuations in returns. We shall also see
that the nature of the two states is quite different.

\subsection*{Learning to trade}

We also consider boundedly rational agents who repeatedly trade in the
market. Each agent $i$ has a propensity to invest $U_i^m(t)$ for each
of the signals $m\in M$. His investment $z_i^m=\chi_i(U_i^m)$ at time
$t$ is an increasing function of $U_i^m(t)$ ($\chi_i: R\to R^+$) with
$\chi_i(x)\to 0$ if $x\to -\infty$ and $\chi_i(x)\to \infty$ if $x\to
\infty$. After each period agents update $U_i^m(t)$ according to the
marginal success of the investment:
\begin{equation}
U_i^m(t+1)=U_i^m(t)+\Gamma
\frac{\partial u_i^\omega}{\partial z_i^m}
\label{learn}
\end{equation}
where $u_i^\omega=\sum_{m\in M}\delta_{k_i^\omega,m}
z_i^m\left(\frac{R^\omega}{p^\omega}-1\right)$ 
is the payoff received in state $\omega$ investing a quantity $z_i^+$ 
or $z_i^-$ depending on $k_i^\omega$
\cite{adaptive}.
As we did with the distinction between competitive equilibrium and Nash equilibrium, we distinguish between naive (or price takers) and sophisticated agents. 
The first do not take into account their impact on the price
and hence take the partial derivative in Eq. (\ref{learn}) {\em at
fixed} $p^\omega$. The latter, instead, account for the fact that
if $z_i^m\to z_i^m+dz$ also the price $p^\omega$ changes by an infinitesimal
amount and this gives a contribution to the partial derivative in 
Eq. (\ref{learn}).

\section{Results}

The study of markets' equilibria and of the asymptotic properties of
the learning dynamics for $N\to\infty$, turns
into the analysis of the minima of the Hamiltonian function
\begin{equation}
H_\eta=|R-p|^2+\eta\frac{\alpha}{N}\sum_{i\in I}\frac{{z_i^+}^2+{z_i^-}^2}{2}
\label{Heta}
\end{equation}
with $\eta=0$ for competitive equilibria or naive agents, and $\eta=1$
for Nash equilibria or sophisticated agents. The first term of
$H_\eta$ (i.e. $H_0$) is just the squared distance between prices and
returns. The competitive equilibrium prices and the Nash equilibrium
prices minimize the Hamiltonian. A detailed proof of this fact shall
be presented elsewhere \cite{long}. Here we offer a brief description
of the main steps leading to it. We also present evidence from
numerical simulations to illustrate our conclusions. First we observe
that, given that $p^\omega-R^\omega = O(1/\sqrt{N})$ is small, one can
linearize the equations in $p^\omega$
\footnote{E.g.  $R^\omega/p^\omega-1\simeq (R^\omega -p^\omega)/\bar R$,
where again $\simeq$ means that we are neglecting terms which vanish at
least as $N^{-1/2}$ relative to the one retained.}.  Finding
equilibria becomes then a linear optimization problem which
can be cast into the minimization of $H_\eta$. 

The learning dynamics can also be linearized, leading to
\begin{equation}
U_i^m(t+1)=U_i^m(t)+\Gamma\delta_{k_i^\omega,m}
\left[(R^\omega-p^\omega)-\eta\frac{z_i^m}{N}\right].
\label{learnlin}
\end{equation}
Here $\eta=0$ describes price takers or naive agents whereas $\eta=1$
describes sophisticated agents. Indeed the term $z_i^m/N$ arises
exactly from a derivative $\frac{\partial p^\omega}{\partial z_i^m}$
of price with respect to investment. The dynamics (\ref{learnlin}), in
the limit $\Gamma\to 0$, turns into\footnote{The idea is to consider a time
interval $\Delta t=d\tau/ \Gamma$ where $d\tau$ is small but much
larger than $\Gamma$. Then $U_i^m(t+\Delta t)-U_i^m(t)$ can be
estimated to linear order in $d\tau$ using the law of large numbers
for $\Delta t\to\infty$ as $\Gamma\to 0$.} a deterministic dynamics
in continuum time. $H_\eta$ is a Lyapunov function of this dynamics
which implies that the asymptotic state is described by the minima of 
$H_\eta$. 

$H_\eta$ is a non-negative definite quadratic form of the dynamical
variables $(z_i^m)_{i\in I, m\in M}$. This means that there is a 
single minimum (which is either a single point or a connected set).

In order to study the properties of the minima of $H_\eta$, we resort to tools of statistical mechanics. $H_\eta$ depends on the particular
realization of the of two random variables: the random information
structure $k_i^\omega$ and of the returns $R^\omega$. In the language of statistical mechanics, these two random variables represent the {\em disorder} in our system. However, in the
limit $N\to\infty$, the statistical behavior is independent of the
specific realization of the disorder\footnote{This statement applies
to some of the observables, which are called self-averaging. Typically
the average value $\sum_i O_i/N$ of a quantity $O_i$ defined for each
agent $i$ is self-averaging (i.e. attain a.s. a fixed value) by virtue
of the law of large numbers when $N\to\infty$. We refer to
ref. \cite{MPV} for further discussion.}. Averages over the disorder
are handled by the {\em replica method} of statistical mechanics
\cite{MPV}, which is briefly discussed in the appendix. A similar
analysis for a system of interacting heterogeneous adaptive agents
has been carried out, with the same techniques, in Ref. \cite{CMZ}.
Below we
discuss the results in the two cases of competitive or Nash equilibria
as a function of the two parameters $\alpha$ and $\sigma$. 

\subsection*{Equilibria and the 
transition to efficient markets}

\begin{figure}
\centerline{\psfig{file=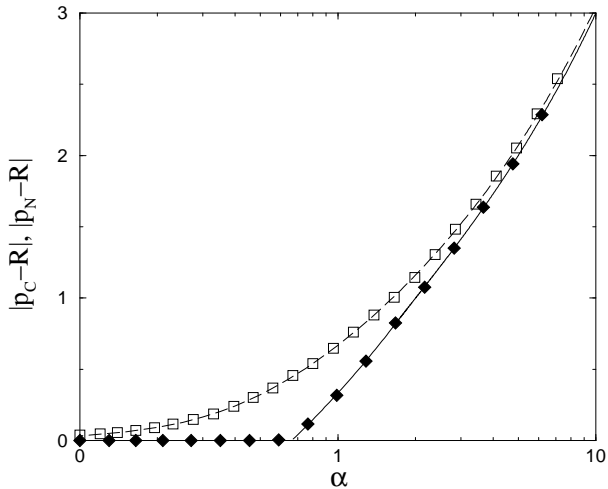,width=8cm}}
\caption{Distance $|p-R|=\sqrt{\sum_\omega(p^\omega-R^\omega)^2}$ of
prices from returns in the competitive (full line and diamonds) and
Nash equilibria (dashed line and squares). Full lines and dashed lines present the analytical graphs; diamonds and squares trace the numerical
simulations of systems with $N=200$ agents and small learning rates.
Averages are taken over $100$ realizations of the disorder in the stationary state.}
\label{figH}
\end{figure}

\begin{figure}
\centerline{\psfig{file=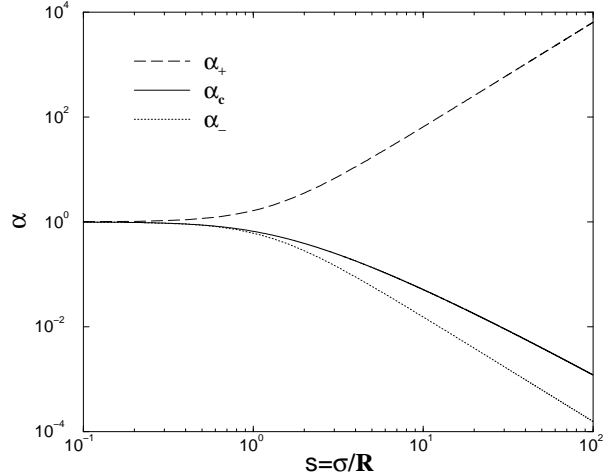,width=8cm}}
\caption{Phase diagram for competitive markets in the $(s,\alpha)$
plane. The curve $\alpha_c(s)$ (full line) separates the efficient
($\alpha<\alpha_c$) from the inefficient ($\alpha>\alpha_c$) phase.
For $\alpha_-<\alpha<\alpha_+$ agents invest at most once, whereas
for $\alpha>\alpha_+$ or $\alpha<\alpha_-$ agents invest a positive
amount under both signals.}
\label{phasediag}
\end{figure}

In the competitive equilibrium agents minimize the distance between prices
and returns. As the number of agents increases, i.e. as
$\alpha=\Omega/N$ decreases, agents are collectively more efficient
in driving prices close to returns. Indeed the distance
$|R-p|$ decreases as $\alpha$ decreases, as shown in
Fig. \ref{figH}. The distance vanishes at a critical point 
$\alpha_c$ which marks a {\em second order phase transition}, 
in the statistical mechanics approach. The value of $\alpha_c$ 
depends on the intensity $\sigma$ of
fluctuations of returns, as shown in Fig. \ref{phasediag}. 
The region
$\alpha<\alpha_c$ is characterized by the condition $H_0=0$, which means
$p^\omega=R^\omega$ for all $\omega\in\Omega$. This means that the
market efficiently aggregates the information dispersed across agents
into the price. The efficient phase, where $H_0=0$, shrinks as
$\sigma$ increases. This is reasonable because as the fluctuations
in $R^\omega$ increase, it becomes harder and harder for the agents 
to incorporate them into prices.

\begin{figure}
\centerline{\psfig{file=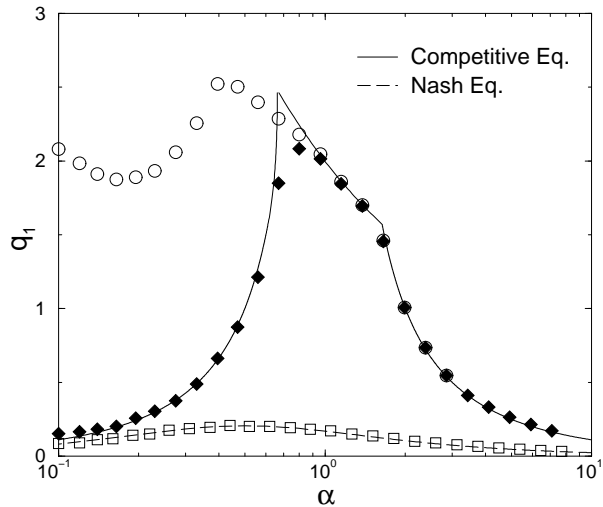,width=8cm}}
\caption{Parameter $q_1$ in the competitive (full line, diamonds and
circles) and Nash equilibria (dashed line and squares). Lines present
the graphs of the analytical solution; diamonds and squares refer to
numerical simulations of systems with $N=200$ agents and small
learning rates.  Averages are taken over $100$ realizations of
disorder in the stationary state. The two sets of simulations, in the
competitive equilibria, refer to different initial conditions:
symmetric (diamonds) and asymmetric (circles).}
\label{figq}
\end{figure}

In order to analyze how agents behave, as a function of $\alpha$ and
$\sigma$, it is useful to introduce the quantity
\begin{equation}
q_1=\frac{1}{N}\sum_{i=1}^N \left(\frac{z^+_i-z_i^-}{2}\right)^2.
\end{equation}
This measures how differently agents behave upon receiving the two
different signals, i.e. how much they use the signal they receive. 
This is plotted in figure \ref{figq} as a function of $\alpha$ for
$\sigma=1$. First we observe that $q_1$ increases as $\alpha_c$ is
approached both from above or below, and that it displays a cusp at
$\alpha_c$. A second singularity appears at a larger value of $\alpha$
-- called $\alpha_+$ -- whose behavior as a function of $\sigma$ is
also shown in Fig. \ref{phasediag}. In the region
$[\alpha_c,\alpha_+)$ agents invest a non-zero amount at most for one 
of the two signals
they receive. Actually in this region there is also a fraction $\phi$
of agents who do not invest ($z_i^+=z_i^-=0$) at all. This fraction
increases as $\alpha_c$ is approached from above, suggesting that 
it becomes harder and harder for agents to find profitable
opportunities with $R^\omega>p^\omega$, as $\alpha\to\alpha_c$.

For $\alpha>\alpha_+$, instead, all agents make positive investments
under both signals ($z_i^+>0$ and $z_i^->0$). This sudden change
manifests itself in the singular behavior of $q_1$ at $\alpha_+$. It is
remarkable that a mixed state, with some agents investing all the
time and some others investing under one signal at most, is not 
possible. As $\alpha$ increases, for $\alpha>\alpha_c$, the 
information complexity increases and the market becomes more and
more inefficient. Prices follow returns to some extent, in the
sense that when $R^\omega$ is larger than the average $-\ovl{R}$, 
the price $p^\omega$ is also larger than $\ovl{p}$. But the slope 
of a regression of $p^\omega-\ovl{p}$ versus $R^\omega-\ovl{R}$
is less than one and it vanishes as $\alpha^{-1}$ for 
$\alpha\to\infty$. 

As the complexity of fluctuations of returns increases, increasing
$\alpha$, the partial information of agents becomes less and less 
useful. Hence for $\alpha\to\infty$ one recovers the case without
information, where $p^\omega=\ovl{R}$, and $H_0\to\alpha\sigma^2$ converges
to the fluctuations of returns themselves.

From the agent's point of view the equilibrium in the efficient phase
($\alpha<\alpha_c$) is not unique. This means that adaptive agents
following the above learning dynamics will end up in a states
$\{z_i^m\}$ which depends on the initial conditions $\{U_i^m(t=0)\}$
(prices, of course, do not depend on the initial condition, because
$p^\omega=R^\omega$ for all $\omega\in \Omega$).
This feature is captured by the parameter
\begin{equation}
q_0=\frac{1}{N}\sum_{i=1}^N\frac{z_i^+-z_i^-}{2}\frac{{z'}_i^+-{z'}_i^-}{2}
\end{equation}
in the statistical mechanics approach, where $z_i^m$ and ${z'}_i^m$
represent two different systems of agents, with different initial
conditions. We find that $q_0\neq q_1$, in general, for
$\alpha<\alpha_c$. Fig. \ref{figq} reports the value of $q_1$ for
symmetric initial conditions $U_i^m(0)=0$ for all $i\in I$ and $m\in
M$. This matches the results of numerical simulations with those
initial conditions but Fig. \ref{figq} also shows that asymmetric
initial conditions ($U_i^m(0)=10m$ for all $i\in I$ and $m=\pm 1$)
yield a different value of $q_1$ and hence a different equilibrium.

\subsection*{Market impact and Nash equilibria}

The fact that the market impact
$\frac{\partial p^\omega}{\partial z_i^m}$ of each agent is of order
$1/N$ might suggest that prices in the Nash equilibrium should be close to those in the competitive equilibrium. More precisely one might expect that for every $\omega$, 
$p_C^\omega-p_N^\omega = O(1/N)$ as in the case where agents have no information on the state of nature that we have discussed above. For large $N$ this is much smaller than
the distance between prices and returns because $p^\omega_C-R^\omega = O(1/\sqrt{N})$ in the competitive equilibrium. This suggests that
Nash equilibria behaves similarly to competitive equilibria, and that the distinction vanishes as $N\to\infty$.

This conclusion however is not correct when agents have private
information, as one can read from Eq. (\ref{Heta}). Indeed, the term
proportional to $\eta$ in Eq. (\ref{Heta}), at the competitive
equilibrium is of the same order in $N$ of the first term.

In fact 
\[
|p_C - p_N| \geq |p_C - R|-|R - p_N|\geq 0
\]
by the reverse triangle inequality ($|p_C - R|-|R - p_N|\geq 0$ holds
because $p_C^\omega$ is the feasible price closest to $R^\omega$). 
Fig. \ref{figH} shows that the right hand side is finite.
This contrast with the case of no information where
$|p_c-p_N|\propto 1/\sqrt{N}\to 0$ as $N\to \infty$. 
This case is only recovered in the limit $\alpha\to\infty$,
in which $|p_C - R|-|R - p_N|\to 0$ (see Fig. \ref{figH}).

The second term of Eq. (\ref{Heta}) dramatically changes the 
statistical behavior of the
market: {\em i)} the phase transition disappears: the distance between
prices $p_N^\omega$ and returns $R^\omega$ smoothly decreases, as
shown in figure \ref{figH}, and it vanishes as $\alpha\to 0$.  {\em
ii)} the equilibrium is unique in both prices and investment for all
$\alpha>0$.  The asymptotic behavior of learning dynamics does not
depend on initial conditions. {\em iii)} Agents always invest at least
under one signal, for all $\alpha>0$. All agents who invest under both
signals invest, on average over the two signals, 
the same quantity\footnote{This is easily
seen from the equilibrium condition, which requires
$\ovl{\delta_{k_i,m}(R-p)}-\ovl{\delta_{k_i,m}}\frac{z_i^m}{N}\le 0$
with the equality holding whenever $z_i^m>0$. Taking the sum on $m$, we find
$\ovl{z_i^{k_i}}/N=\ovl{(R-p)}$ for all agents who invest under both
signals. The same conditions imply that there cannot be an agent who
never invests. Indeed if $z_i^m=0$, it must be that
$\ovl{\delta_{k_i,m}(R-p)}\le 0$. The sum on $m$ gives $\ovl{(R-p)}\le
0$ which cannot be correct: agents will never spend more that they get
at equilibrium.}. {\em iv)} average price $\ovl{p}$ 
is lower than average return $\ovl{R}$, by a term of order $1/N$.
As a consequence, the payoff of agents is not 
zero, as in competitive equilibria, but rather it is proportional
to $1/N$.

Again as $\alpha\to\infty$ the fluctuations of prices with $\omega$
die out. This explains why, in that limit the price converges to the
one without information. At the same time, as $\alpha\to\infty$
$|p_N-R|-|p_C-R|\to 0$ and Nash prices 
converge to competitive prices (see Fig. \ref{figH}). 

\subsection*{Learning rates and price volatility}

\begin{figure}
\centerline{\psfig{file=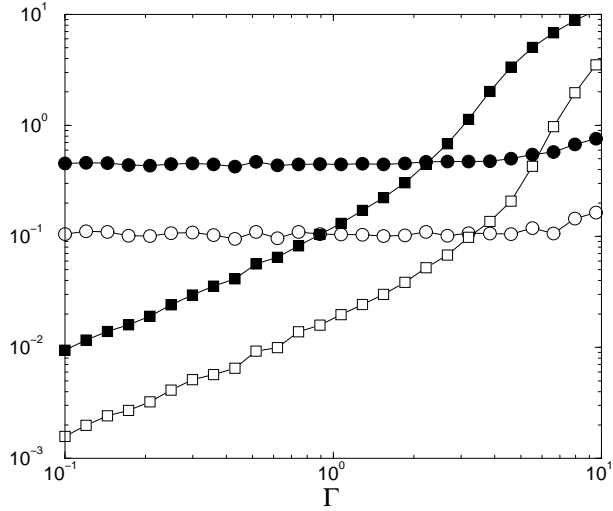,width=8cm}}
\caption{Squared distance $|R-\avg{p}|^2$ of prices from returns
for $\alpha=s=1$ as a function of $\Gamma$ (circles). Price 
fluctuation $\delta p^2$ are also plotted (squares). Open symbols
refer to Competitive equilibria whereas full symbols refer to
Nash equilibria.}
\label{dpCENE}
\end{figure}

Let us explore the dependence of the results discussed so far on the
learning rate $\Gamma$. In the limit $\Gamma\to 0$ the dynamics is
deterministic and hence prices converge to equilibrium prices and the
investment of agents also converges to a fixed point equilibrium.
When $\Gamma>0$ the dynamics becomes stochastic and fluctuations
occur.  These do not affect average prices $\avg{p|\omega}$, which
attain their $\Gamma\to 0$ value\footnote{The notation $\avg{\ldots}$
is intended here for long time averages in the stationary state of the
dynamics.  The symbol $\avg{\ldots|\omega}$ stands averages
conditional on the state $\omega$.}. On the contrary, price volatility
increases, as shown in Fig. \ref{dpCENE}. For both Competitive and
Nash equilibria we find that the distance of average prices to returns
$|R-\avg{p}|$ stays remarkably constant when $\Gamma$ increases over
two decades. On the contrary, price fluctuations $\delta p^2\equiv
\sum_\omega\avg{(p-\avg{p|\omega})^2|\omega}/\Omega$ increase
approximately linearly with $\Gamma$ in the same range.

Hence excess volatility arises in these model markets when agents react
too fast or strongly to price adjustments. 

\section{Conclusions}

We have characterized the relative size of the market that gives at
equilibrium market efficiency. The economy we have considered is very
simple: there is only one asset. In addition, to simplify the
analysis, we have considered the case of agents with unbounded wealth.

Our results generalize in a straightforward way to more general
setups. This is true in particular for the economies with many
assets. As long as agents have no budget constraint, they can invest
independently in each asset separately. In other words agents do not
introduce additional correlation in the prices of different assets
beyond those already present in the returns.

\appendix

\section{Appendix: the statistical mechanics analysis}

In order to study the minima of $H_\eta$, we introduce the partition function
\[
Z(\beta)=\int_0^\infty dz_1^+\int_0^\infty dz_1^-\ldots
\int_0^\infty dz_N^+\int_0^\infty dz_N^- e^{-\beta H_\eta\{z_i^m\}}.
\]
The integrals, in the limit $\beta\to \infty$ are dominated by the
configurations $\{z_i^m\}$ for which $H_\eta$ is minimal. The central
quantity to compute is the {\em free energy} $f_\beta=-\beta^{-1} \log
Z(\beta)$. This has to be averaged over the realization of the {\em
disorder} $\{k_i^\omega, R^\omega\}$. One can reduce the problem of
computing the average $\avg{\log Z(\beta)}$ of the logarithm to that
of computing averages of moments $\avg{ Z^n(\beta)}$ by the identity
$\log Z=\lim_{n\to 0} (Z^n-1)/n$.  This is known as the replica trick
\cite{MPV} because, for integer $n$, $Z^n(\beta)$ is the partition
function of $n$ non-interacting replicas
$\{z_{i,1}^m\},\ldots,\{z_{i,n}^m\}$ of the system, with the same
realization of disorder. Taking the average over the disorder
introduces an effective interaction between replicas. The resulting
expression is handled in such a way as to be able to use saddle point
methods to compute it in the limit $N,\beta\to\infty$.  An assumption
on the symmetry of the minima with respect to permutation of replicas
is necessary at this point. This symmetry can be broken in case of
multiple minima, because different replicas may end-up in different
minima. This is not our case because $H_\eta$ is a non-negative
definite quadratic form. So the solution in our case is {\em replica
symmetric} (see \cite{MPV} for more details). A more detailed
discussion of these steps in the context of a model of adaptive
heterogeneous agents is given in Ref. \cite{CMZ}.

The final result is that we can write
\begin{equation}
E_{k,R}\left[\min_{\{z_i^m\}} H_\eta\right]=
\lim_{\beta\to\infty} \min_{q_0,\hat q_0,x,w,\hat R} 
f_{\beta}(q_0,\hat q_0,x,w,\hat R)
\end{equation}	
where 
\bea
f_{\beta}(q_0,\hat q_0,x,w,\hat R)&=&\frac{1}{\beta}\ln(1+x)
+\frac{q_0+\sigma^2}{1+x}+\frac{x\hat q_0}{\alpha}-\frac{w q_0}{\alpha}
+\frac{2\hat R}{\alpha}\nonumber\\
&~&-\frac{2}{\alpha\beta}\int_{-\infty}^\infty 
\!\frac{dt\,e^{-t^2/2}}{\sqrt{2\pi}}
\ln \int_{z\ge 0}
\!d^2 z e^{-\beta V_t(z)}
\nonumber
\eea
and, with $z=(\bar z+\Delta,\bar z -\Delta)$
\be
V_t(z)=\frac{w+\alpha\eta}{2}
\Delta^2-\sqrt{\hat q_0} t\Delta 
+\frac{\alpha\eta}{2}{\bar z}^2-\hat R\bar z.
\ee
Here $x=\beta(q_1-q_0)$ is a combination of the two order
parameters whereas $\hat q_0,w$ and $\hat R$ arise as auxiliary 
variables (Lagrange multipliers). 

The minimization problem contains the statistical information 
on the Gibbs probability distribution $e^{-\beta H_\eta}$ over
the space spanned by $\{z_i^m\}$, averaged over all realizations
of disorder. We refer the reader to ref. \cite{MPV} for a deeper
discussion. Let it suffice to say that this probability distribution
factorizes over agents, which is typical of systems with mean field
interaction. Hence the probability that an agent invests fractions
$z\in A$ where $A$ is any subset of $R^2_+$, is given by
\be
{\rm Prob}\{z\in A\}=\int_{A}\!dz
\int_{-\infty}^\infty 
\!\frac{dt\,e^{-t^2/2}}{\sqrt{2\pi}}\frac{e^{-\beta V_t(z^+,z^-)}}
{\int_{\zeta\ge 0}d^2\zeta e^{-\beta V_t(\zeta)}}
\label{Pz}
\ee
where the parameters of $V_t$ are those which satisfy 
the five first order conditions
$\frac{\partial f_\beta}{\partial q_0}=0$, 
$\frac{\partial f_\beta}{\partial \hat q_0}=0$, 
$\frac{\partial f_\beta}{\partial x}=0$, 
$\frac{\partial f_\beta}{\partial w}=0$ and 
$\frac{\partial f_\beta}{\partial \hat R}=0$.

As $\beta\to\infty$, the integrals in Eq. (\ref{Pz}) are dominated
by the minimum $z^*$ of $V_t(z)$ in $R^2$. The calculation leading
to the final result are lengthy. We just report the final result.

\subsection{The case of $\eta=0$} 

When $\eta=0$ a degeneracy of solutions occurs for $\alpha<\alpha_c$.
In other words, $H_0$ does not attain its minimum at a single isolated
point, but rather on a connected set of points. Each point is the
equilibrium of a market with some initial condition. As one varies 
the initial conditions, the equilibrium moves on the set. 

The statistical mechanics approach takes an uniform average over all
the points of this set. The results derived in this way, do not
describe any particular market equilibrium, but rather a uniform 
distribution
of equilibria resulting from some distribution of initial conditions.
Since the mapping between initial conditions and equilibria is not
known, it is not clear what exactly the statistical mechanics approach
describes. 

There is way out, in these kind of situations, which is to introduce a
term in the functional to be minimized which breaks the degeneracy and
selects one point in the set. In the limit when the strength of this
term vanishes, one recovers the statistical properties of the original
system in the selected equilibrium. The $\eta$ term is exactly of this
form. Furthermore it is a perturbation which preserves the symmetries
of the system ($k_i\to -k_i$ for some $i$). With $\eta=0^+$ we then
expect to find the properties of markets with symmetric initial
conditions ($U_i^+(0)=U_i^-(0)$). 

The solution for $\eta=0^+$ takes a parametric form. Let us define the 
functions 
\beas
\psi_r(\tau)&=&
\sqrt{\frac{2}{\pi}}e^{-\tau^2/2}-\tau {\rm erfc}(\tau/\sqrt{2})\\
\psi_q(\tau)&=&
(1+\tau^2) {\rm erfc}(\tau/\sqrt{2})-\sqrt{\frac{2}{\pi}}\tau e^{-\tau^2/2}\\
\psi_x(\tau)&=&{\rm erfc}(\tau/\sqrt{2}).
\eeas

For $\tau\in [0,\tau_c]$, where $\tau_c$ is the solution of
$\psi_q(\tau)+\sigma^2\psi_r(\tau)=\psi_x(\tau)$ we have

\beas
\alpha&=&\psi_q(\tau)+\sigma^2\psi_r(\tau)\\
q_0&=&\frac{\psi_q(\tau)}{\psi_r^2(\tau)}\\
|R-p|&=&\frac{\psi_q(\tau)+\sigma^2\psi_r(\tau)-\psi_x(\tau)}
{\psi_r(\tau)}
\eeas

which holds in the interval 
$\alpha\in [\alpha_c,1+\frac{2}{\pi}\sigma^2]$, where 
$\alpha_c=\alpha(\tau_c)$. 
Notice that $H\to 0$ as $\alpha\to\alpha_c$.

For $\alpha\in [(1+\frac{2}{\pi}\sigma^2)^{-1},\alpha_c]$ we find
\beas
\alpha&=&\frac{\psi_x(\tau)}{\psi_q(\tau)+\sigma^2\psi_r(\tau)}\\
q_0&=&\frac{\psi_q(\tau)}{\psi_r^2(\tau)}\\
|R-p|&=&0
\eeas
again for $\tau\in [0,\tau_c]$. 

In these two regions, a fraction $\phi={\rm erf}(\tau/\sqrt{2})$
of agents never invests, whereas the others invest only under 
at most one signal. The fraction $\phi$ is largest at the critical
point $\alpha_c$, i.e. for $\tau=\tau_c$.

Outside this region we find an explicit solution: For $\alpha\ge
1+\frac{2}{\pi}\sigma^2$, we find $q_0=1/\alpha$ and
$|R-p|=(1-1/\alpha)\sqrt{1+\alpha\sigma^2}$. Hence the bare 
fluctuations of returns $|R-1|=\sqrt{\alpha\sigma^2}$ are recovered 
in this limit. In the region
$\alpha\le 1/(1+\frac{2}{\pi}\sigma^2)$ we find $q_0=\alpha$ and
$|R-p|=0$. Agents invest under both signals and the distribution of
their average investment $(z^++z^-)/2$ is exponential.

\subsection{The case of $\eta=1$}

The study of the minimum of the potential $V_t(z)$ is quite different
in this case. One has to distinguish two regions of integration of $t$,
according to whether the minimum of $V_t(z)$ belongs to $R^2_+$ or not.
The first case describes agents who invest under both signals, and 
our solution confirms that they all invest on average the same amount. 
The second case describes the fraction of agents who only invest once.

The solution again depends parametrically on $\tau$ and on a parameter
$y\in[1,2]$ which satisfies the equation:
\beas
1-\frac{I_0(\tau)}{1 + y}&=&
  \frac{( 2 - y ) ( y-1 )}{y}
   \left\{ 1-\frac{I_0(\tau) + 
        [ I_1(\tau) - \tau]\tau + 
        2I_0(\tau)y}{{( 1 + y) }^2}\right.\\
      &~&~~~~~~~~~~~~~~~~~~~~~~~+\left. \frac{\sigma^2 
        {\left[\tau + ( I_1(\tau) + \tau - 
               I_0(\tau)\tau)y \right] }^2}
        {{( 1 + y ) }^2} \right\}
\eeas
with $I_0(\tau)={\rm erfc}(\tau/\sqrt{2})$ and 
$I_1(\tau)=\sqrt{2/\pi} e^{-\tau^2/2}$. We have
\beas
q_0&=&\frac{(1+y)^2+\tau^2-(1+2y)I_0(\tau)-\tau I_1(\tau)}
{[\tau(1+y)-y\tau I_0(\tau)+y I_1(\tau)]}\\
\alpha&=&(q_0+\sigma^2)
\left\{\frac{(y-1){[\tau(1+y)-y\tau I_0(\tau)+y I_1(\tau)]}}{y(1+y)}
\right\}^2\\
|R-p|&=&(q_0+\sigma^2)(y-1)^2\left[\frac{\tau}{y}-
\frac{\tau I_0(\tau)-I_1(\tau)}{1+y}\right]
\eeas

This has two solutions $y^\pm$ for $\tau>\tau_0$, which describe the
low and high $\alpha$ regions. There is no singularity or discontinuity
in the solution when $\tau\to\tau_0$ (corresponding to $\alpha\approx
0.502\ldots$ for $\sigma^2=1$). The fraction of agents playing only 
once is given by $I_0(\tau)$.

\end{document}